\documentclass[epj]{svjour}
\usepackage{graphics}

\begin{document}

\newcommand{\so}
   {\mathrel{\rlap{\raise1pt\hbox{$>$}}{\lower4pt\hbox{$\sim$}}}}
\newcommand{\io}
   {\mathrel{\rlap{\raise1pt\hbox{$<$}}{\lower4pt\hbox{$\sim$}}}}

\title{N-body Study of Anisotropic Membrane Inclusions: Membrane
  Mediated Interactions and Ordered Aggregation}

\titlerunning{Anisotropic inclusions}

\author{P. G. {\sc Dommersnes} \and J.-B. {\sc Fournier}}
\institute{Laboratoire de Physico-Chimie Th{\'e}orique,
  E.\,S.\,P.\,C.\,I.\,, 10 rue Vauquelin, F-75231 Paris C{\'e}dex 05,
  France.}
\date{Received: date / Revised version: date}
\abstract{
  We study the collective behavior of inclusions inducing local {\em
    anisotropic\/} curvatures in a flexible fluid membrane. The
  $N$-body interaction energy for general anisotropic inclusions is
  calculated explicitly, including multi-body interactions. Long-range
  attractive interactions between inclusions are found to be
  sufficiently strong to induce aggregation. Monte Carlo simulations
  show a transition from compact clusters to aggregation on lines or
  circles.  These results might be relevant to proteins in biological
  membranes or colloidal particles bound to surfactant membranes.
\PACS{
      {87.15.Kg}{}   \and
      {64.60.Cn}{}   \and
      {24.10.Cn}{}
     }
}
\maketitle

The interplay between structural features and $N$-body interactions is a
general physical problem arising in many different contexts, e.g.,
crystals structure~\cite{Lotrich}, magnetic atom
clusters~\cite{Pastor,PastorII}, colloids in charged fluids~\cite{Netz},
polyelectrolyte condensation~\cite{HaLiu,PodPar}, and protein
aggregation in biological membranes~\cite{Oster}. $N$-body interactions
can sometimes yield spectacular effects: non-pairwise summability
of charge fluctuation forces can dramatically affect the stability of
polyelectrolyte bundles~\cite{HaLiu}; three-body elastic
interactions may induce aggregation of membrane inclusions, although
two-body elastic interactions are repulsive~\cite{Oster}. In a system
able to kinetically achieve equilibrium, the clusters formed are
usually compact, however certain interactions may favor tenuous
clusters.  For instance, a recent $N$-body study has shown that above
a critical strength of three-body interactions, the state of minimum
energy is one in which all the particles are on a line~\cite{Date}. It
has also been observed recently that membrane mediated interactions
can induce one-dimensional ring-like aggregates of colloidal particles
bound to fluid vesicle membranes~\cite{Safinya}.

Manifolds embedded in a correlated elastic medium can impose boundary
conditions, or modify the elastic constants. This usually gives rise
to mean-field forces, which are due to the elastic deformation of the
medium, and to {\em Casimir\/} forces, which are due to the
modification of its thermal fluctuations. Such interactions are
generally non pairwise additive~\cite{LiKardarII}. The elastic
interactions between defects in solids~\cite{Landau} or in liquid
crystals~\cite{deGennes} are well known examples of mean-field forces.
Casimir forces exist between manifolds embedded in correlated fluids,
such as liquid crystals and
superfluids~\cite{Armand,LiKardarI,LiKardarII}, or critical
mixtures~\cite{Fisher}. Another interesting example is the interaction
between inclusions in flexible membranes~\cite{Gil98}: it has been
shown that cone shaped membrane inclusions experience both long range
attractive Casimir interactions and repulsive elastic interactions
falling of as $R^{-4}$ with separation~$R$~\cite{Goulian93}.

In this Rapid Note, following Netz~\cite{Netz97}, we give exact
results concerning the long range multi-body interactions among
membrane inclusions that break the bilayer's up-down symmetry.
However, rather than supposing that the inclusions simply induce a
local spontaneous curvature, we assume that the inclusions {\em set\/}
a preferred curvature tensor~\cite{Goulian93,Park96}. This model is
more realistic: the ``preference'' of a conically shaped inclusion is
$c_1=c_2=c_0$, where $c_1$ and $c_2$ are the membrane principal
curvatures, rather than the weaker condition $c_1+c_2=2c_0$ assumed in
Ref.~\cite{Netz97}. In addition, the imposed curvature tensor can be
anisotropic, thus describing inclusions that break the in-plane
symmetry. In a first part we calculate the exact Casimir and
mean-field two- and three-body interactions between such anisotropic
inclusions. Then, the {\em collective\/} behavior of identical
inclusions is investigated by means of a Monte Carlo (MC)
simulation, using the full $N$-body interaction energy plus a
hard-core repulsion modeling the simplest repulsive short-range
interactions~\cite{Dan94}.  Our results could be relevant to
understanding the aggregation and organization of proteins in
biological membranes, or colloidal particles bound to surfactant
membranes~\cite{Safinya}.

Let us consider a system of $N$ anisotropic inclusions embedded in a
flexible fluid membrane, in which they are free to diffuse laterally.
In many situations, the surface tension is negligible and the membrane
shape is governed by the Helfrich curvature energy
$h_0=\frac{1}{2}\kappa\,(c_1+c_2)^2 +\bar{\kappa}\,c_1c_2$~\cite{Helfrich73},
where $\kappa$ is the bending rigidity, and $\bar{\kappa}$ the Gaussian modulus.
For biological membranes, $\kappa\sim30\,T$, while for surfactant membranes it
can be as small as a few $T$ ($T$ will denote throughout the
temperature in energy units). We model the membrane shape by a simple
parametric surface $({\bf r},u({\bf r}))$, where ${\bf r}$ is a vector
in the $(x,y)$ plane and $u({\bf r})$ the normal displacement field
along $z$. To quadratic order in $u$, the Helfrich Hamiltonian takes
the form
\begin{equation}
  \label{eq:Helfrich}
  {\cal H}_0=\int\!d{\bf r}\left[\frac{1}{2}\kappa\left(\nabla^2 u\right)^2
+\bar\kappa\,{\rm det}\left(\nabla\nabla u\right)\right].
\end{equation}
Its correlation function, $\langle u({\bf 0})u({\bf
  r})\rangle=(T/\kappa)\,G(\bf{r})$ is given by the Green function
\begin{equation}
  G({\bf r})=\left(\nabla^4\right)^{-1}\!\delta({\bf r})=
  G(0)-\frac{1}{8\pi }\,r^2\ln\!\left(\frac{L}{r}\right),
\end{equation}
where $L$ is a long wavelength cut-off, which is comparable with the
size of the membrane.

Typical membrane inclusions have a central hydrophobic region spanning
the hydrophobic core of the membrane, and two polar extremities
protruding outside~\cite{BioBook}. Assuming a strong coupling between
the lipids and the inclusion's boundary~\cite{Owicki79,Dan94}, which
in general is not cylindrical, we model membrane inclusions as
curvature sour\-ces~\cite{Goulian93}, {\em point-like\/} as
in~\cite{Park96}.  Indeed, the size of the proteins is comparable with
the short wavelength cut-off, i.e., the membrane thickness. Even large
particles can be treated as point-like inclusions, provided one uses
the curvature energy coarse-grained to the size of inclusions with its
corresponding renormalized bending rigidity~\cite{Helfrich85,Peliti}.
Hence, for an inclusion located at ${\bf r}_n$, we enforce the local
condition
\begin{equation}
\left.\nabla\nabla u\right|_{{\bf r}_n}={\bf Q}_n\equiv\left( 
\begin{array}{c}
   K_n-J_n  \\ 0
\end{array}
\begin{array}{c}
   0 \\ K_n+J_n
\end{array}
     \right),
\end{equation}
where $\nabla\nabla u$ is the curvature tensor of the membrane, formed
by the second derivatives of $u$, and ${\bf Q}_n$ is the curvature
tensor locally imposed by the inclusion. Here, ${\bf Q}_n$ is written
in the frame of reference where it is diagonal, $K_n$ being the
mean-curvature and $J_n$ the anisotropic curvature of the inclusion.
While isotropic inclusions correspond to $J_n\!=\!0$, ``hyperbolic''
and wedge-shaped inclusions correspond to $K_n\!=\!0$ and
$K_n\!=\!J_n$, respectively.  In the following, we may assume
$J_n\geq0$ without loss of generality.  We define the {\em
  orientation\/} of inclusion $n$ by the direction of the lowest
principal curvature, i.e., $K_n-J_n$. The latter is defined modulo a
rotation of angle $\pi$. Indeed, even if the inclusion bears an
in-plane polarity, its coupling with the membrane curvature is apolar
by symmetry.
 
Within this model, we can calculate the free energy $\cal F$ of a
membrane with $N$ inclusions in a non perturbative way. Introducing
an external field $h({\bf r})$, eventually set to zero, the free 
energy is given by
\begin{equation}
  \exp{[-\frac{\cal F}{T}]} = \int \tilde{{\cal D}}[u]
               \exp\left[-\frac{1}{T}\!\int\!\! d{\bf r}\;
u\!\left(\kappa\nabla^4\right)\!u+h\,u\right]\!, 
\end{equation}
in which the constraints are implemented by delta functions in
the measure
\begin{equation}
            \tilde{{\cal D}}[u] = {\cal D}[u] 
            \prod_{n=1}^{N}\delta
\left(\nabla\nabla u|_{{\bf r}_n}-{\bf Q}_n\right).
\end{equation}
This integral will be regularized by introducing a microscopic
cutoff~$a$ of the size of the membrane thickness. The Gaussian
curvature energy, which yields a constant contribution depending on
the boundary conditions at the edge of the membrane and a constant
self-energy per inclusion $\simeq\!-\pi a^2\bar\kappa(K_n^2-J_n^2)$
(consistent with Ref.~\cite{Goulian93}), has been discarded.

Replacing the delta functions by their Fourier representation (for
details on this method, see, e.g., Ref.~\cite{LiKardarII}), and
integrating out the field $u({\bf r})$, we obtain
 \begin{eqnarray}
\label{energy}
{\cal F}[h]&=&\frac{T}{2}\ln\left(\det{\bf M}\right)+
\frac{1}{2}\kappa\,{\bf b'}\,{\bf M}^{-1}\,{\bf b'}^{\rm t},\\
{\bf b'}&=&{\bf b}+\!\int\!\!d{\bf r}\,h({\bf r})\,{\bf c}({\bf r}).
\end{eqnarray}
${\bf M}$ is a $3N\times 3N$ matrix formed by the blocks ${\bf
  M}_{np}={\bf D}^{\rm t}{\bf D}\,G({\bf r}_n-{\bf r}_p)$, where
${\bf D}=(\partial^2_{x},\partial_x\partial_y,\partial^2_{y})$; 
${\bf b}$ and ${\bf c}({\bf r})$  are $3N$ vectors with blocks ${\bf
  b}_n=[({\bf Q}_n)_{11},({\bf Q}_n)_{12},({\bf Q}_n)_{22}]$ and
${\bf c}_n={\bf D}\,G({\bf r}-{\bf r}_n)$, respectively.

\begin{figure}
  \resizebox{0.5\textwidth}{!}
  {\hspace*{3cm}\includegraphics{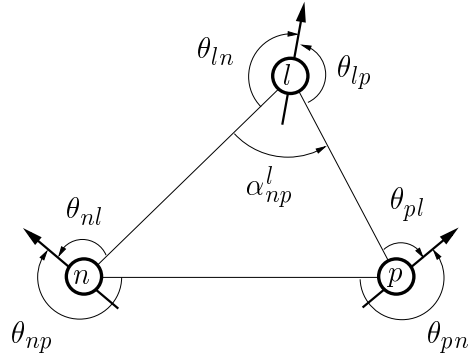}\hspace*{3cm}}
  \caption{Definition of the mutual orientations of 
    three inclusions labeled $n,m$, and $l$. The angle $\theta_{nl}$
    is the orientation of inclusion $n$ with respect to the line
    joining inclusions $n$ and $l$, etc. All angles are defined modulo
    $\pi$, since any orientation and its reverse are equivalent.}
  \label{Fig1}
\end{figure}

The average shape of the membrane is given by 
\begin{equation}
\label{shape}
\langle u({\bf r}) \rangle=
\left. \frac{\delta {\cal F}[h]}{\delta h({\bf r})}\right |_{h=0}
=\,{\bf c}({\bf r})\,{\bf M}^{-1}\,{\bf b}^{\rm t}.
\end{equation}
Although not in explicit form, Eq.~(\ref{energy}) allows to calculate
{\em exactly\/} the $N$-body interaction, by inverting the matrix ${\bf M}$
whose elements are simple functions of the distances between
particles.  ${\bf M}$ and ${\bf b}$ do not depend on $T$, therefore the
first term in~(\ref{energy}) corresponds to the fluctuation induced
Casimir interaction, while the second term represents the
mean-field elastic interaction.

Expanding the Casimir interaction ${\cal F}^{\rm C}\!=\!
\frac{1}{2}T\ln(\det{\bf M})$ to
fourth order in $1/r_{np}$, where $r_{np}=|{\bf r}_n-{\bf r}_p|$,
yields
\begin{equation}
 {\cal F}^{\rm C}_4=-3T{\sum_{n,p}}'\frac{a^4}{r_{np}^4},
\end{equation}
in which the prime indicates restriction of the sum to different
inclusions. This fourth-order Casimir interaction is pairwise additive
and corresponds exactly to the sum of the two-body interactions
obtained in Refs.~\cite{Goulian93,Park96} for isotropic inclusions. It
does not depend on the locally imposed curvature nor on the
orientation of the inclusions (unlike the Casimir interaction between
two rod-like inclusions~\cite{Gol}). This interaction is quite weak,
since for $r>2a$ it is only a fraction of $T$ and thus it is
overwhelmed by diffusion.

Setting $h=0$ in the second term of~(\ref{energy}) yields the
mean-field elastic energy ${\cal F}^{\rm mf}=\frac{1}{2}\kappa\,{\bf
  b}\,{\bf M}^{-1}{\bf b}^{\rm t}$, which arises from the bending
deformation of the membrane. To second order in $1/r_{np}$, we obtain
(see Fig.~\ref{Fig1} for the definition of the angles)
\begin{eqnarray}
\label{F2}
 {\cal F}_2^{\rm mf} =&& -4\pi\kappa{\sum_{n,p}}' \frac{a^4}{r_{np}^2}
\big[2J_nJ_p\cos(2\theta_{pn}\!-\!2\theta_{np})
\nonumber\\
  &&+K_nJ_p\cos2\theta_{pn}+K_pJ_n\cos2\theta_{np}\big].
\end{eqnarray}
At this order, the mean-field interaction is pairwise additive.  It
depends on the imposed curvature tensors and on the orientations of
the principal curvatures. To {\em fourth\/} order, the mean-field
interaction includes both two-body interactions:
\begin{eqnarray}
\label{F4a}
{\cal F}_{42}^{\rm mf}&&=\!2\pi\kappa {\sum_{n,p}}'
\frac{a^6}{r_{np}^4}
\big[(5+\cos4\theta_{np})J_n^2
+(5+\cos4\theta_{pn})J_p^2\nonumber\\
+&&K_n^2\!+\!K_p^2\!+\!4K_nJ_n\cos2\theta_{np}
  \!+\!4K_pJ_p\cos2\theta_{pn}\big],
\end{eqnarray}
and {\em three-body\/} interactions:
\begin{eqnarray}
\label{F4b}
  {\cal F}&&_{43}^{\rm mf}=
\frac{4\pi}{3}\kappa{\sum_{n,p,\ell}\!}'
\frac{a^6}{r_{n\ell }^2r_{\ell p}^2}\Big\{\!
J_nJ_p\big[4\cos(2\theta_{n\ell }\!+\!2\theta_{p\ell }
\!+\!2\alpha_{np}^{\ell})\nonumber\\
+&&2\cos2\theta_{n\ell}\cos2\theta_{p\ell}\big]
+2K_nJ_p\cos(2\theta_{p\ell }\!+\!2\alpha_{np}^{\ell})
\nonumber\\
+&&2K_pJ_n\cos(2\theta_{n\ell }\!+\!2\alpha_{np}^{\ell})
\!+\!\!K_nK_p\cos2 \alpha_{np}^{\ell}\!\Big\}\!+{\rm perm.}
\end{eqnarray}
These terms can be attractive or repulsive,
depending on the curvatures and orientations of the inclusions.
Setting $J_n\!=\!J_p\!=\!0$ in~(\ref{F4a}), we recover the result of
Refs.~\cite{Goulian93,Park96} for two isotropic inclusions; however,
we also notice that the three-body interactions are in general of the same
order as the two-body interactions.

Let us study the lowest-order mean-field interaction ${\cal F}_2^{\rm
  mf}$ between two identical anisotropic inclusions. Setting
$K_1\!=\!K_2\!=\!K$ and $J_1\!=\!J_2\!=\!J$, we may assume $K\geq 0$
and $J\geq 0$ without loss of generality. If $K\neq0$, the energy is
minimal when $\theta_{12}=\theta_{21}=0$. The inclusions therefore
tend to align their axis of smallest principal curvature (smallest in
modulus) parallel to their separation vector. If $K=0$, the energy is
minimal whenever $\theta_{12}=\theta_{21}$. In both cases the
interaction is attractive, contrary to the repulsive interaction found
between two isotropic inclusions~\cite{Goulian93,Park96}. If
$\theta_{12}=0$ and $\theta_{21}\equiv\theta$ is arbitrary, the
interaction is repulsive for
$\theta^{\star}<|\theta|<\pi-\theta^{\star}$, where
$\cos{2\theta^{\star}}=-K/(K+2J)$, and otherwise attractive; the
relation $\pi/4<\theta^{\star}<\pi/2$ holds, where the lower limit
corresponds to $K=0$, and the upper limit to $J\to 0$. When
$\theta_{12}=\theta_{21}\equiv\theta$, the interaction is also
repulsive for $\theta_c<|\theta|<\pi-\theta_c$ and attractive
otherwise, with $\cos{2\theta_c}=-J/K$. Again, we have
$\pi/4<\theta_c<\pi/2$, where the lower limit corresponds to $J
\to 0$ and the upper limit to $J=K$ (wedge-shaped inclusions).
If $J>K$ the interaction is always attractive when
$\theta_{12}=\theta_{21}$.

\begin{figure}
  \resizebox{0.5\textwidth}{!}
  {\hspace*{1cm}\includegraphics{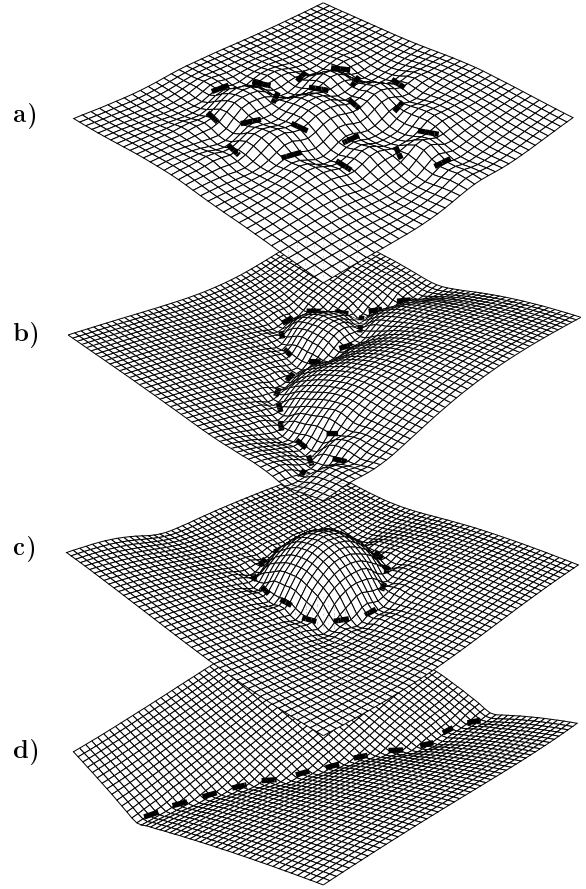}\hspace*{1cm}}
  \caption{Typical equilibrium aggregates obtained from MC simulation
    of $N=20$ identical inclusions. The corresponding values of $(\bar
    K,\bar J)$ are indicated in the phase diagram of Fig.~\ref{Fig3}.
    The mesh size corresponds to the microscopic cut-off $a$. The bars
    indicate the orientations of the inclusions.}
  \label{Fig2}
\end{figure}

As seen above, the lowest-order interaction between two identical
anisotropic inclusions is already quite complex, however it is
essentially attractive.  Hence, one expects aggregation of the
particles when the interaction energy overcomes the entropy of mixing.
We noticed that the case $K=0$ is special, since the minimum energy is
degenerate.  Thus, one might expect compact clusters in this case, but
for larger values of $K$ it should become increasingly favorable to
orient the inclusions along a common line joining them.  To check
these ideas, we have performed MC simulations on finite samples of
$N\!=\!20$ identical inclusions. The positions and directions of
principal curvatures were varied continuously, in order to avoid
lattice artifacts. The energy was calculated from the $N$-body
interaction (\ref{energy}), with $h=0$, by numerically inverting the
$3N\times3N$ matrix ${\bf M}$.  When inclusions are close to contact,
additional short-range interactions should be taken into
account~\cite{Gil98,Dan94,Owicki79,JB}. To mimic them, we have added a
{\em repulsive\/} hard-core potential (diameter $4a$), in agreement
with the results of Ref.~\cite{Dan94} in which a short-range energy
barrier was obtained from a microscopic model.  Obviously, in the case
of attractive short-range interactions, we cannot predict the
aggregate shapes within the present model. The results of our MC
studies merely give the aggregation {\em tendencies\/} resulting from
the long-range interactions.

\begin{figure}
  \resizebox{0.5\textwidth}{!}
  {\hspace*{2cm}\includegraphics{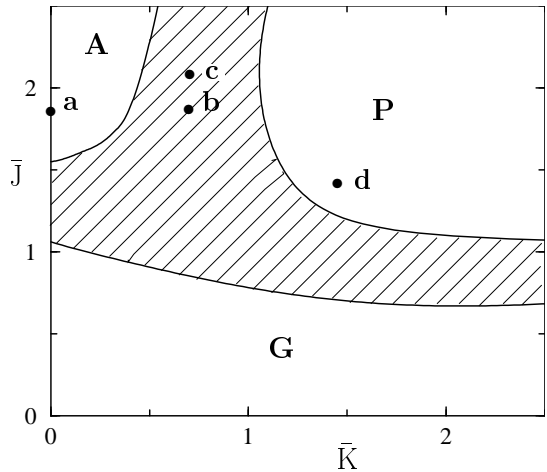}\hspace*{2cm}}
  \caption{Phase diagram of identical anisotropic inclusions obtained by 
   MC simulations for different values of the inclusions
   mean-curvature $\bar K$ and anisotropic curvature $\bar J$. (G) gas
   phase, (A) compact aggregates, and (P) linear polymer-like
   aggregates. In the dashed region, we found branched
   and ring-like aggregates, or small clusters close to the
   gas phase.}
  \label{Fig3}
\end{figure}

We started the simulations with a low-density distribution of
inclusions having both random positions and orientations, then we let
the system equilibrate using $\sim\!10^7$ MC steps. The relevant
dimensionless parameters are ${\bar K}=(\kappa/T)^{1/2} Ka$ and
$\bar{J}=(\kappa/T)^{1/2} Ja$. For weak anisotropy, i.e., $\bar{J} \io 1$,
the interaction was either repulsive or too weakly attractive to
induce aggregation; for $\bar{J}\so1$, the interaction was
sufficiently attractive for the inclusions to aggregate. For small
$\bar{K}$, we obtained compact clusters of inclusions, inducing some
kind of decorated ``egg-carton'' structure of the membrane
(Fig.~\ref{Fig2}a).  As $\bar{K}$ increased, we found a transition
from compact aggregates to polymer-like ones (Fig.~\ref{Fig2}d).
Remarkably, we also found ring-like aggregates, featuring a budding
mechanism (Fig.~\ref{Fig2}b,c). The boundaries between the different
aggregates shapes are depicted in the phase diagram of
Fig.~\ref{Fig3}. Performing simulations with $N=60$, we observed
relatively weak fluctuations of the linear aggregates, indicating a
persistence length of several hundred of inclusions or more. This
suggest that ``polymers'' made of wedge-shaped inclusions can acquire
a persistence length in the micron range. Long-range order might exist
in the form of parallel inclusion lines, since for a two dimensional
object, the $1/r^2$ interaction is marginally long-range. In all these
aggregates, the multi-body interactions are of the same order as the
pairwise interactions, but the Casimir interaction is not determinant
for the types of aggregates obtained.

Our MC simulations concerned only systems of {\em identical\/}
inclusions, however they may easily be extended to the case of
inclusions inducing different curvatures. It would also be interesting
to study the effect of the long-range anisotropic interactions in the
situation of kinetically constrained aggregation.  In the case of a
film with surface tension, we found similar anisotropic interactions,
but of shorter range $\sim\!1/r^4$ at lowest order~\cite{article}.
   
We thank R. Bruinsma and J. Prost for useful discussions. P. G. D. was
supported by the Research Council of Norway.

\end{document}